\documentclass[twocolumn,twoside,slac]{revtex4}
\usepackage{graphicx}
\usepackage{fancyhdr}

\begin{document}

\title{Worldwide Fast File Replication on Grid Datafarm}

\author{Osamu Tatebe, Satoshi Sekiguchi}
\affiliation{AIST, Tsukuba, Ibaraki 3058568, JAPAN}
\author{Youhei Morita}
\affiliation{KEK, Tsukuba, Ibaraki 3050801, JAPAN}
\author{Satoshi Matsuoka}
\affiliation{Tokyo Institute of Technology, Meguro, Tokyo 152-8552, JAPAN}
\author{Noriyuki Soda}
\affiliation{Software Research Associates, Inc., Naka, Nagoya, 4600003, JAPAN}

\begin{abstract}
The Grid Datafarm architecture is designed for global petascale
data-intensive computing.  It provides a global parallel filesystem
with online petascale storage, scalable I/O bandwidth, and scalable
parallel processing, and it can exploit local I/O in a grid of
clusters with tens of thousands of nodes.  One of features is that it
manages file replicas in filesystem metadata for fault tolerance and
load balancing.

This paper discusses and evaluates several techniques to support
long-distance fast file replication.  The Grid Datafarm manages a
ranked group of files as a Gfarm file, each file, called a Gfarm file
fragment, being stored on a filesystem node, or replicated on several
filesystem nodes.  Each Gfarm file fragment is replicated
independently and in parallel using rate-controlled HighSpeed TCP with
network striping.  On a US-Japan testbed with 10,000 km distance, we
achieve 419 Mbps using 2 nodes on each side, and 741 Mbps using 4
nodes out of 893 Mbps with two transpacific networks.
\end{abstract}

\maketitle

\thispagestyle{fancy}

\section{Introduction}
\label{sec:intro}
Petascale data intensive computing wave has been coming such as
high-energy physics data analysis, astronomical data analysis, and
bio-informatics data analysis.  More than ten petabyte storage needs
to be shared and analyzed by world-wide users with high efficiency,
high security, and high dependability.

The Grid Datafarm architecture \cite{gfarm-ccgrid2002} is designed for
global petascale data-intensive computing to enable the process of
large amounts of data at multiple regional PC clusters.  The aim of
this research is to establish a large-scale parallel filesystem by
exploiting local storage of cluster nodes spread in the extensive
area, a platform system needed to support a petabyte scale data
intensive computing.  The Grid Datafarm architecture enables
high-speed access to a large amount of data by utilizing file access
locality, and realizes fault tolerance of disks and networks by data
replication.

This paper discusses about long-distance fast file replication on the
Grid Datafarm.  To improve the performance in high bandwidth-delay
product networks, congestion control keeping efficient, fair,
scalable, and stable plays a key role.  The easiest way to improve the
performance is to open multiple TCP connections in parallel, while
this approach leaves the parameter of the number of connections to be
determined by the user, which may result in heavy congestion with too
much number of connections.  There are several researches addressing
this issue such as HighSpeed TCP \cite{HighSpeedTCP}, Scalable TCP
\cite{ScalableTCP}, FAST TCP \cite{FASTTCP}, and XCP
\cite{XCP-SIGCOMM2002}.  HighSpeed TCP is an attempt to improve
congestion control of TCP for large congestion windows with better
flexibility, better scaling, better slow-start behavior, and competing
more fairly with current TCP, keeping backward compatibility and
incremental deployment.  It modifies the TCP response function only
for large congestion windows to reach high bandwidth reasonably
quickly when in slow-start, and to reach high bandwidth without overly
long delays when recovering from multiple retransmit timeouts, or when
ramping-up from a period with small congestion windows.

For file replication of large files in high bandwidth-delay product
networks, it is also necessary to improve disk I/O performance not
only the network performance.  At this time, each cluster node has
capability to transmit at a rate of 1 Gbps, while the performance of
an IDE or a SCSI disk is at most 50 MB/s.  To improve the disk I/O
bandwidth, disk striping such as RAID-0 is effective.

This paper is organized as follows.  In Section~\ref{sec:rep}, the
file replication on the Grid Datafarm is discussed.
Section~\ref{sec:eval} evaluates the network performance and the file
replication performance using a US-Japan Grid Datafarm testbed.


\section{File Replication on Grid Datafarm}
\label{sec:rep}
The Grid Datafarm provides a Grid file system that federates multiple
local filesystems on a Grid across administrative domains.  The Grid
file system provides virtualized hierarchical namespaces for files
having consistent access control with flexible capabilities
management.  There is a replica catalog to manage mappings from the
hierarchical namespace for files to one or more physical file
locations.  This enables efficient, dependable, and transparent file
sharing on a Grid.

The Grid Datafarm has a feature for data parallel execution.  It
manages a ranked group of files as a single Gfarm file.  This makes it
possible to manage a lot of distributed files as a single file, which
will be analyzed in parallel.  Each parallel process possibly
generates a new set of output files also managed as a single file.
File-affinity scheduling and new concept of a file view enable the
``owner computes'' strategy, or ``move the computation to data''
approach for the parallel data analysis.

When replicating a Gfarm file, each file of the Gfarm file, or a group
of files, stored on a different filesystem node can be replicated in
parallel and independently.  File replication on the Grid Datafarm is
considered to be a parallel file replication from multiple cluster
nodes to (different) multiple cluster nodes.

In high bandwidth-delay product networks, multiple TCP streams, or
network striping, is effective to improve the performance, while disk
accesses to striping data decreases the performance of the disk I/O\@.
It is necessary to utilize a modified TCP or other protocols to
achieve high performance with a single stream.  HighSpeed TCP is one
of proposals of the modification of congestion control of TCP, which
is utilized by the performance evaluation on a US-Japan Grid Datafarm
testbed.

As described in Section~\ref{sec:intro}, the disk I/O performance is
poorer than the network bandwidth on each cluster node.  One of
requirements from the Grid Datafarm architecture is a dense and
high-performance storage on each node for online petabyte-scale
storage.  To meet this requirement, each node of the AIST Gfarm
cluster was designed to be a 1U server having a 3ware RAID card with
four 120GB IDE disks in RAID 0 that achieves 480 GB storage capacity
and over 110 MB/s for contiguous block reads and writes that is
comparable with the network performance.

\section{Performance Evaluation}
\label{sec:eval}

\subsection{PC clusters and wide-area networks}
\label{ssec:env}
During the international conference SC2002, held in Baltimore from
November 16th to November 22nd of 2002, a Grid Datafarm Data Grid
environment was set up linking seven PC clusters in both Japan and the
U.S.

\begin{figure}[tb]
 \includegraphics[width=\columnwidth]{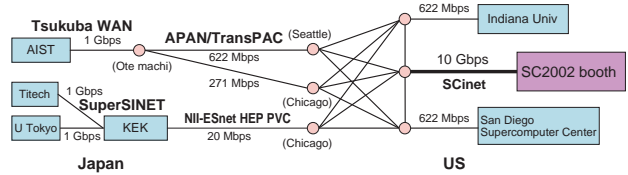}
 \caption{\label{fig:env} Network logical map of world-wide Grid
 Datafarm testbed.  Four sites in Japan and three sites in U.S. are
 integrated with the Grid Datafarm Data Grid middleware.}
\end{figure}

Seven systems of PC clusters, including the one at the SC2002 booth (a
total of 190 PCs) were located in the research centers of both Japan
and the U.S. (AIST, High Energy Accelerator Research Organization,
Tokyo Institute of Technology, the University of Tokyo, Indiana
University and San Diego Supercomputer Center (SDSC)).  They were
integrated with the Grid Datafarm Data Grid middleware
\cite{gfarm-home} (Figure~\ref{fig:env}).  Tsukuba WAN, APAN/TransPAC
\cite{apan-transpac} and Maffin \cite{maffin} supported in
establishing the network.

The system had the peak floating-point performance of 962 Gflops.  It
was equipped with a large capacity file system of 18 TB at 6,600 MB/s
access rate as a Gfarm wide-area filesystem.

This environment utilized multiple high-speed wide-area networks, that
is, Tsukuba WAN and SuperSINET in Japan, APAN/TransPAC and NII-ESnet
HEP PVC between Japan and the U.S., Abilene and ESnet in the U.S. and
SCinet at the SC2002 booth.  The bandwidth from SC2002 to both Indiana
University and SDSC was 622 Mbps.  In the transpacific network, it was
893 Mbps.  Total theoretical maximum bandwidth of the network linking
seven clusters was 2.173 Gbps one way (See Figure~\ref{fig:env}).

For the file replication, a large amount of scientific data taken from
particle physics was generated mainly in the large-scale PC cluster of
Tokyo Institutes of Technology and created data replicas of several
hundreds of GB at each of the other clusters in a single filesystem
image.

At the SC2002 booth in Baltimore, there was a 12-node AIST Gfarm
cluster connected with gigabit ethernet that connects to the SCinet
with 10 gigabit ethernet using the Force10 E1200 switch.  Each node
consisted of a dual Intel Xeon 2.4GHz processor, 1GB memory, and a
3ware Escalade 7500-4 RAID controller with four 120GB 3.5'' HDDs,
which was configured in RAID-0.  The disk I/O performance for
contiguous blocks achieved 109 MB/s for writes and 168 MB/s for reads.
The network bandwidth of gigabit ethernet was 941 Mbps using the iperf
bandwidth measurement tool \cite{iperf}.  File replication performance
was 75 MB/s, that was equivalent to 629 Mbps, using the Gfarm Data
Grid Middleware.


At the Grid Technology Research Center, AIST in Tsukuba, Japan, there
was the same 7-node AIST Gfarm cluster that connects to the Tokyo XP
with gigabit ethernet via Tsukuba WAN and Maffin networks.

At the Indiana University, there was a 15-node PC cluster connected
with Fast Ethernet connects to Indianapolis GigaPoP with OC-12.  The
disk I/O performance for contiguous blocks was 9.3 MB/s for writes and
10.2 MB/s for reads.


At the SDSC in San Diego, there was a 8-node PC cluster connected with
gigabit ethernet that connects to outside with OC-12.  The disk I/O
performance for contiguous blocks was 29.4 MB/s for writes and 20.0
for reads.


APAN/TransPAC transpacific network consisted of two links; the
northern route (OC-12 POS) between Seattle and Tokyo, and the southern
route (OC-12 ATM) between Chicago and Tokyo.  The southern route was
shaped to 271 Mbps.  By default, all IP packets were transmitted via
the northern route.  To utilize the both routes, we configured a
static route such that every traffic between specific three nodes at
SC2002 booth and AIST was transmitted via the southern route.

\begin{table}[tb]
\caption{Round trip time between the SC2002 booth and other sites.
'AIST (N)' and 'AIST (S)' mean that IP packets are transmitted via the
northern route and via the southern route, respectively}
\label{tab:rtt}
\begin{center}
\begin{tabular}{cc}
\hline
AIST (N) & 199 msec \\
AIST (S) & 222 msec \\
Indiana & 30 msec \\
SDSC & 86 msec \\
\hline
\end{tabular}
\end{center}
\end{table}

The round trip time (RTT) of IP packets between the SC2002 booth and
other sites is shown in the Table~\ref{tab:rtt}.

\subsection{HighSpeed TCP over transpacific network}
\begin{figure}[tb]
 \includegraphics[width=\columnwidth]{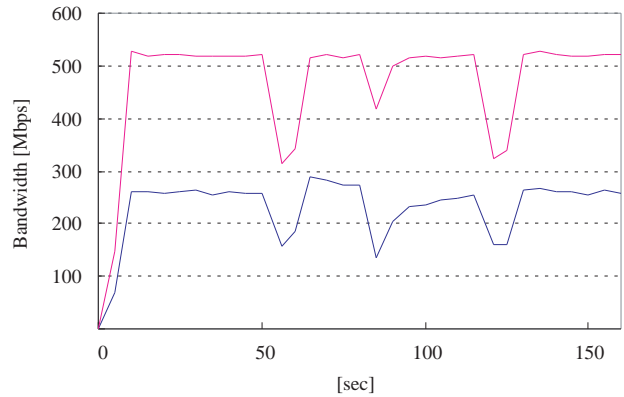}
 \caption{\label{fig:sc-gfm} HighSpeed TCP bandwidth from U.S. to
 Japan via the APAN/TransPAC northern route (OS-12 POS).}
\end{figure}

Figure~\ref{fig:sc-gfm} shows the measured network bandwidth of the
HighSpeed TCP from the SC2002 booth in U.S. to the AIST in Japan via
the APAN/TransPAC northern route.  The bandwidth was measured using
the iperf from one node to one node with two HighSpeed TCP streams.
The buffer size of each socket was 8 MB, which gave the theoretical
peak bandwidth 337 Mbps for one connection with the RTT 199 ms.  From
the Figure~\ref{fig:sc-gfm}, the measured peak bandwidth achieved 529
Mbps in 5-second average out of the physical network bandwidth 622
Mbps.  Due to the packet loss, the bandwidth occasionally dropped,
however, it was recovered reasonably quickly thanks to the HighSpeed
TCP\@.


\begin{figure}[tb]
 \includegraphics[width=\columnwidth]{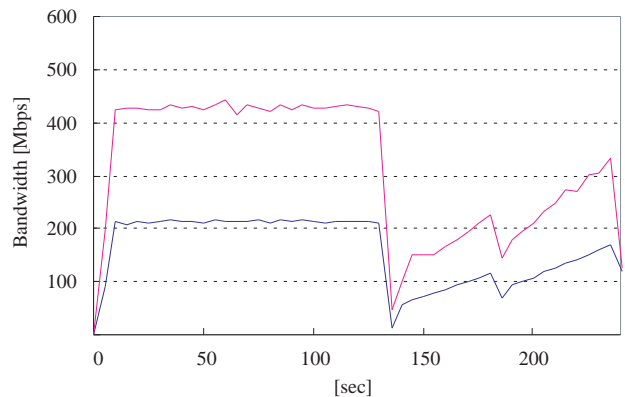}
 \caption{\label{fig:gfm-sc} HighSpeed TCP bandwidth from Japan to
 U.S.  via the APAN/TransPAC northern route (OS-12 POS)}
\end{figure}

Figure~\ref{fig:gfm-sc} shows the measured HighSpeed TCP bandwidth
from Japan to U.S. via the northern route.  The machine and network
configurations were the same as the previous measurement except the
traffic direction.  Because the traffic was slightly heavy at this
time, the measured peak bandwidth was 443 Mbps.  After the heavy
packet loss, the bandwidth was recovered slowly just like the regular
TCP\@.

\begin{figure}[tb]
 \includegraphics[width=\columnwidth]{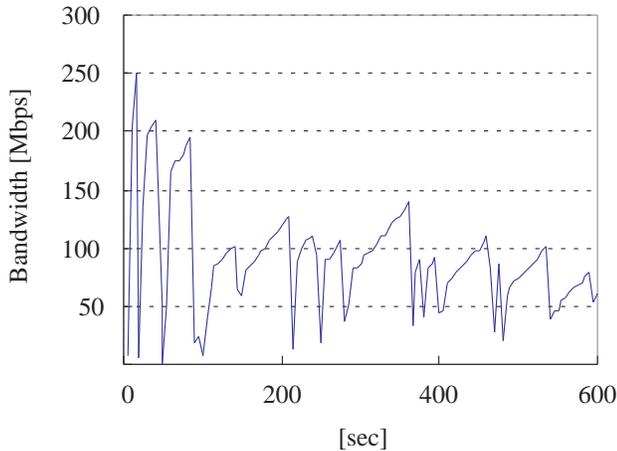}
 \caption{\label{fig:aist-sdsc-ge} HighSpeed TCP bandwidth from Japan
 to U.S. via the APAN/TransPAC southern route (OC-12 ATM, 271 Mbps
 shaping) with one non-rate-limited HighSpeed TCP stream.}
\end{figure}

Figure~\ref{fig:aist-sdsc-ge} is the case using the APAN/TransPAC
southern route.  Since the southern route was shaped to 271 Mbps, one
HighSpeed TCP stream would be able to fill the bandwidth.  However,
the stream suffered the critical packet loss, and only achieved 85.9
Mbps in 10-minutes average although 251 Mbps in 5-second peak
bandwidth.  One of reasons for the critical packet loss was the
setting of an ATM switch configured without the random early drop.  To
cope with this problem, it is necessary to control the network traffic
rate not to excess the physical network bandwidth, that is, 271 Mbps.

\begin{figure}[tb]
 \includegraphics[width=\columnwidth]{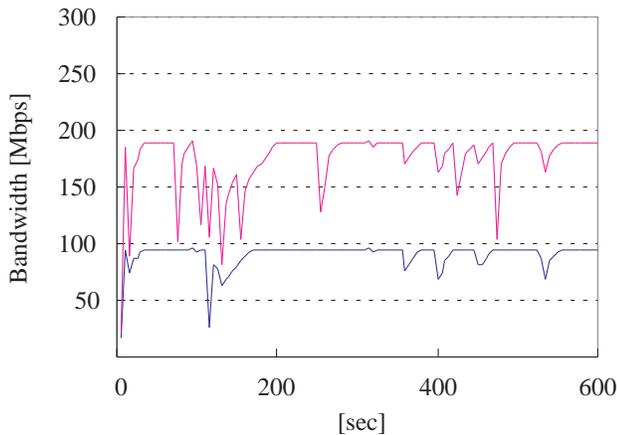}
 \caption{\label{fig:aist-sdsc-fe2} HighSpeed TCP bandwidth from Japan
 to U.S. via the APAN/TransPAC southern route (OC-12 ATM, 271 Mbps
 shaping) with two 100-Mbps HighSpeed TCP streams.}
\end{figure}

When the maximum bandwidth of one HighSpeed TCP stream was limited to
100 Mbps, it was possible to achieve stable and high bandwidth shown
by Figure~\ref{fig:aist-sdsc-fe2}.  This rate control was realized by
changing the network from gigabit ethernet to fast ethernet.  This
case achieved 166.1 Mbps in 10-minutes average, and 190.0 Mbps in
5-second peak bandwidth.

\subsection{File replication}
The performance of file replication of large files that do not fit the
main memory is limited by the disk I/O performance and the network
performance.  When replicating between sites, the file replication
performance is also limited by the bandwidth of the wide-area network
shown by Figure~\ref{fig:env}.  Table~\ref{tab:max-rep} shows the
performance limit of file replication with one node at each site.

\begin{table}[tb]
\begin{center}
\caption{Performance limit of file replication using one node at each
site in MB/s.}
\label{tab:max-rep}
\begin{tabular}{c|cc}
        &   To   &   From \\
\hline
Indiana &   9.3  &   10.2 \\
SDSC    &  29.4  &   20.0 \\
AIST    & 109    &  112   \\
SC2002  & 109    &  112   \\
\end{tabular}
\end{center}
\end{table}

\begin{figure}[tb]
 \includegraphics[width=\columnwidth]{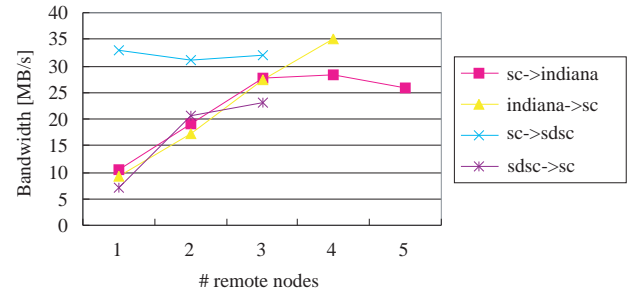}
 \caption{\label{fig:sc-ussites} File replication performance between
 one node at SC2002 booth and several nodes at Indiana Univ.\ and SDSC\@.}
\end{figure}

Figure~\ref{fig:sc-ussites} shows the performance of file replication
between one node at the SC2002 booth and various number of nodes at
Indiana Univ.\ and SDSC\@.  Between the SC2002 booth and Indiana
Univ., the file replication performance increased almost proportional
to the number of nodes at Indiana Univ., and achieved the maximum
performance of 34.9 MB/s, that is, 293 Mbps, from four nodes at
Indiana Univ.\ to one node at the SC2002 booth with a 4GB file.

From the SC2002 booth to SDSC, the file replication performance of a
2GB file achieved 32.8 MB/s even with one node at SDSC\@.  On the
other hand, the performance from SDSC to the SC2002 booth showed a
different tendency to increase scalable with respect to the number of
nodes at SDSC\@.  One reason was this direction requires multiple
HighSpeed TCP streams to improve the network performance because one
stream achieved only 1.1 to 2.3 Mbps.  For the file replication from
SDSC to the SC2002 booth, seven parallel streams were used in each
node pair.

Table~\ref{tab:us-japan-rep} showed parameters and the measured
bandwidth of file replication from Baltimore in U.S. to Tsukuba in
Japan at a distance of more than 10,000 km.  As shown in the previous
section, a HighSpeed TCP stream achieved about 260 Mbps with socket
buffer size 8 MB, while multiple rate-controlled streams were
effective to stabilize the bandwidth.

\begin{table*}[th]
\begin{center}
\caption{Parameters and measured performance of US-Japan file replication}
\label{tab:us-japan-rep}
\begin{tabular}{cccccc}
\# node pairs & \# streams & data size & 10-sec average BW
 & Transfer time & Average BW \\
\hline
1 (N1)    &  1 (N1x1)  & 2 GB & n/a & 162.6 sec & 106 Mbps \\
1 (N1)    &  8 (N8x1)  & 2 GB & n/a & 124.7 sec & 138 Mbps \\
1 (N1)    & 16 (N16x1) & 2 GB & n/a  & 113.0 sec & 152 Mbps \\
\hline
1 (S1)    &  1 (S1x1)  & 1 GB & n/a   & 193.0 sec & 44.5 Mbps \\
1 (S1)    &  8 (S8x1)  & 1 GB & 170 Mbps &  91.5 sec & 93.9 Mbps \\
1 (S1)    & 16 (S16x1) & 1 GB & n/a   & 173.3 sec & 49.6 Mbps \\
\hline
2 (N2)    & 32 (N16x2) & 2$\times$2 GB & 419 Mbps & 115.9 sec & 297 Mbps \\
3 (N3)    & 48 (N16x3) & 2$\times$3 GB & 593 Mbps & 139.6 sec & 369 Mbps \\
4 (N3+S1) & 56 (N16x3+S8x1) & 2$\times$4 GB & 741 Mbps & 150.0 sec & 458 Mbps \\
\end{tabular}
\end{center}
\end{table*}

To control and limit the traffic at any rate, the socket buffer size
and the interval of sending data were adjusted.  The interval of
sending data was also needed to be adjusted to suppress too fast
increase of the congestion window that causes the serious packet loss.

Using the northern route, 16 streams achieved the best bandwidth of
152 Mbps in average for file replication of a 2 GB file.  Using the
southern route, 8 streams achieved the best bandwidth of 93.9 Mbps in
average for file replication of a 1 GB file.  Using three node pairs
for the northern route and one node pair for the southern route, the
file replication of a 8 GB file achieved 458 Mbps in average, and 741
Mbps in 10-second peak bandwidth out of 893 Mbps.


For the SC2002 high-performance bandwidth challenge, parameters of
Table~\ref{tab:sc2002-bwc} was set up based on the previous
measurement.  In the remote site column, `AIST N' means the AIST via
the APAN/TransPAC northern route, and `AIST S' means the AIST via the
southern route.  The `Measured BW' column shows the measured average
bandwidth of file replication of a 1 GB or 2 GB file, which is not the
same as 5-second or 10-second peak bandwidth.

\begin{table*}[tb]
\begin{center}
\caption{Parameters for file replication and expected bandwidth from
and to a 12-node Gfarm cluster at SC2002 booth}
\label{tab:sc2002-bwc}
\begin{tabular}{cccccc}
\multicolumn{6}{c}{Outgoing traffic} \\
\# nodes & Remote & \# nodes & \# streams &
Socket buffer size, & Measured BW \\
in Baltimore & site & at remote site & /node &
rate limit & (1-2min avg) \\
\hline
 3 &  SDSC    & 5 &  1 & 1 MB              & $>$ 60 MB/s \\
 2 &  Indiana & 8 &  1 & 1 MB              &  56.8 MB/s \\
 3 &  AIST N  & 3 & 16 & 610 KB, 50 Mbps   &  44.0 MB/s \\
 1 &  AIST S  & 1 & 16 & 346 KB, 28.5 Mbps &  10.6 MB/s \\
\hline
 9 & S,I,A    & 5,8,4 & - & & $>$ 171 Mbps \\
   &          &       &   & & ($>$ 1.43 Gbps) \\
\multicolumn{6}{c}{Incoming traffic} \\
\hline
 1  &  SDSC    & 3 &  7 & 7 MB             &  23.1 MB/s \\
 1* &  Indiana & 4 &  1 & 1 MB             &  34.9 MB/s \\
 1* &  AIST N  & 1 &  1 & 610 KB, 50 Mbps  &   n/a \\
 1  &  AIST S  & 1 &  1 & 346 KB           &   n/a \\
\hline
 3  & S,I,A    & 3,4,2 & - & & $>$ 58 MB/s \\
    &          &       &   & & ($>$ 487 Mbps)
\end{tabular}
\end{center}
\end{table*}

The average bandwidth in one to two minutes would be expected to be
achieved over 1.43 Gbps for outgoing traffic and over 487 Mbps for
incoming traffic, over 1.92 Gbps in total for both directions, if
there were no unknown congestion and no unexpected packet drop by
using several networks simultaneously.

\begin{figure}[tb]
 \includegraphics[width=\columnwidth]{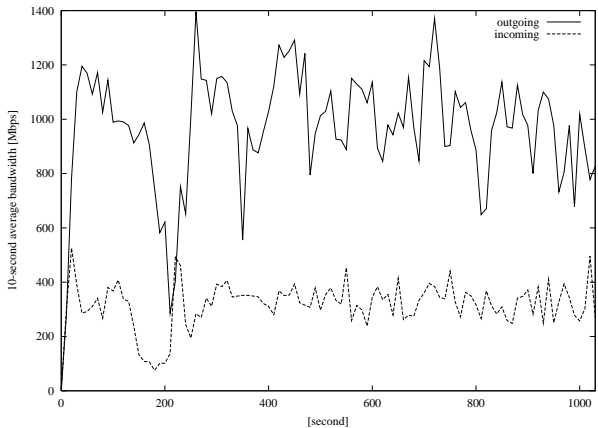}
 \caption{\label{fig:bwc02} File replication performance in 10-second
 average between the SC2002 booth and other sites during the SC2002
 high-performance bandwidth challenge.}
\end{figure}

As a result, the file replication performance was shown by
Figure~\ref{fig:bwc02} in 10-second average bandwidth.  The peak
bandwidth was 1.40 Gbps for outgoing traffic, and 0.526 Gbps for
incoming traffic.  The 0.1-second average bandwidth measured by the
SCinet showed 1.691 Gbps for outgoing traffic, and 0.595 Gbps for
incoming traffic, 2.286 Gbps in total for both directions using 12
nodes in Baltimore.


\section{Summary and Future Work}
The Grid Datafarm is an architecture for petabyte-scale data-intensive
computing providing online ten petabyte-scale storage, an I/O
bandwidth scales to the TB/s range and scalable computational power,
which is securely and dependably shared on a Grid.  This paper
discussed and evaluated the performance of file replication on the
Grid Datafarm.

For the evaluation of the network performance in high bandwidth-delay
product networks between U.S. and Japan, the HighSpeed TCP performed
very well on the transpacific network of OC-12 POS, and achieved 529
Mbps in 5-second average using two streams of one node pair.  On the
other hand, the application-level rate-control of a HighSpeed TCP
stream was necessary for the network of OC-12 ATM to achieve stable
and high bandwidth.

Within the U.S., the file replication showed any performance problem,
while between U.S. and Japan, application-level rate-control of a
HighSpeed TCP were also needed for stability.  As a result, using
three node pairs for the northern route and one node pair for the
southern route, the file replication of a 8 GB file achieved 741 Mbps
in 10-second average out of 893 Mbps.

Between the SC2002 booth and other three sites including a Japan site,
the peak bandwidth of file replication in 0.1-second average showed
1.691 Gbps for outgoing traffic, and 0.595 Gbps for incoming traffic,
2.286 Gbps in total for both directions using 12 nodes in the SC2002
booth.

The Grid Datafarm can be applied to theoretical or experimental
science that calls upon large-scale data analysis and simulation.  We
are planning to evaluate it using large-scale production applications
such as high-energy physics data analysis, analysis of observational
data of all-sky multiple wavelength bands in astronomy, gene analysis
in bio-informatics and so on on a world-wide Grid Datafarm testbed.

\section*{Acknowledgments}
We would like to thank kind help for PRAGMA members, especially, Rick
McMullen, John Hicks at Indiana Univ., Phillip Papadopoulos at SDSC\@.
We are thankful to the web100 and net100 projects for providing a
HighSpeed TCP patch for a linux kernel.  We are grateful to Hisashi
Eguchi at Maffin, Kazunori Konishi, Yoshinori Kitatsuji, and Ayumu
Kubota at APAN, Chris Robb at Indiana Univ.\ for investigation of
bottlenecks of wide-area networks.  We appreciate great help of Force
10 Networks, Inc.\ for providing the E1200 switch with 10 gigabit
ethernet network interface.  This research was supported by the
Ministry of Economy, Trade and Industry through research grant of
Network computing project, and the Ministry of Education, Culture,
Sports, Science and Technology of Japan through a Grant-in-Aid for
Scientific Research on Priority Areas (2) (No.\ 13224034).


\end{document}